# Relaxation dynamics of glasses along a wide stability and temperature range

C. Rodríguez-Tinoco, J. Ràfols-Ribé, M. González-Silveira, J. Rodríguez-Viejo*

Physics Department, Universitat Autònoma de Barcelona, 08193 Bellaterra, Spain.

**Abstract**

While lot of measurements describe the relaxation dynamics of the liquid state, experimental data of the glass dynamics at high temperatures are much scarcer. We use ultrafast scanning calorimetry to expand the timescales of the glass to much shorter values than previously achieved. Our data show that the relaxation time of glasses follows a super-Arrhenius behaviour in the high-temperature regime above the conventional devitrification temperature heating at 10 K/min. The liquid and glass states can be described by a common VFT-like expression that solely depends on temperature and limiting fictive temperature. We apply this common description to nearly-isotropic glasses of indomethacin, toluene and to recent data on metallic glasses. We also show that the dynamics of indomethacin glasses obey density scaling laws originally derived for the liquid. This work provides a strong connection between the dynamics of the equilibrium supercooled liquid and non-equilibrium glassy states.

*Correpondence to javier.rodriguez@uab.cat

One of the biggest challenges in condensed matter physics is the understanding of amorphous systems, which lack the long range order of crystalline materials[1–5]. In spite of it, glasses are ubiquitous in our day life and many materials with technological significance display disordered atomic or molecular arrangements[1]. Amorphous solids are usually obtained from the liquid state avoiding crystallisation. The relaxation time of the liquid increases exponentially during cooling, at a pace determined by its fragile or strong nature. In the laboratory time scale, around certain value of the relaxation time, the molecules do not have enough time to explore the complete configurational space and get trapped inside local energy minima, forming a glass[1–5]. Below this temperature, upon further cooling, the relaxation time of the glass follows a much softer Arrhenius-like expression[6]. For many years, there has been an increased interest in the time scales of physical processes occurring below the glass transition temperature, $T_g$, due to the importance of understanding and controlling relaxation processes in the glass. On the other hand, the inherent unstable nature of glasses has prevented detailed investigations of their properties during heating at temperatures above the conventional $T_g$, where a glass would irreversibly relax into the equilibrium liquid state.

Several models have been developed to comprehend the supercooled liquid (SCL) dynamics and the glass transition phenomena. Among them, the Adam-Gibbs (AG) formalism has provided a suggestive connection between the dynamics and the thermodynamics of amorphous systems[7]. This model has been able to describe the relaxation behaviour of deeply supercooled liquids remarkably well, yielding the well-known Vogel-Fulcher-Tamman, VFT, equation[8], which is often used to evaluate the dynamics of supercooled liquids,

$$\tau_\alpha = \tau_0 e^{\frac{DT_0}{(T-T_0)}} \qquad (1)$$

where $\tau_0$ is the limiting value of $\tau$ at an infinite temperature, D is a material constant related to its fragility and $T_0$ is the diverging temperature. Many other theories are invoked to extend the modelling to the behaviour of liquids and glasses, such as the random first-order transition theory (RFOT)[2], the potential energy landscape (PEL)[3], the mode-coupling theory (MCT)[9], or the Coupling Model (CM)[10]. The relaxation time of glasses has generated certain debate in the glass science community[6,11]. Much below $T_g$, in the glass state, the configurational entropy of the system remains constant and, therefore, it is generally accepted that the dependence of the glass relaxation time with temperature responds to an Arrhenius expression[6]. However, due to the intrinsically slow relaxation times of such systems below the glass transition temperature, the access to experimental data requires enormous amounts of time, which makes measurements impractically long[12]. On the other hand, at higher temperatures, the glass irreversibly transforms into the supercooled liquid in shorter time scales. In this range, the access to relaxation time values requires both ultrafast heating and a rapid dynamic response, accessible through fast scanning nanocalorimetry[13–16]. The influence of stability on the relaxation time of the glass is also a relevant topic in the current literature[17,18]. A new procedure to increase the stability of a glass is to grow it by vapour-deposition at temperatures around 0.85 $T_g$[19,20]. In optimum conditions, the stability of vapour-deposited glasses can be equivalent to the stability of conventional glasses aged for thousands or millions of years or cooled at rates many orders of magnitude slower than conventional methods allow[21]. Therefore, by tuning the deposition conditions, vapour-deposited glasses offer a convenient route to explore the influence of stability on the melting of the glass over a much larger range than ever before.

Here, we perform heat capacity measurements in a broad range of heating rates, from 0.167 K/s up to $2 \cdot 10^4$ K/s, of indomethacin (IMC) and toluene glasses embedded with different kinetic and thermodynamic stabilities. We also fit recent experimental data by Wang et al.[22] on Au-based metallic glasses to support our conclusions. With the high heating rates achieved with fast scanning calorimetry, we expand the accessible timescales of the glass to much lower values than currently reported, which permits us to infer the dynamics over a large temperature interval. We propose that the kinetic behaviour of a liquid and all its isotropic glasses respond to the same dependence with the temperature and the thermodynamic stability of the system, evaluated through its enthalpic limiting fictive temperature. We also show that glasses of different stability, and therefore with different density, fulfil density scaling relations[23–25] that were originally derived for the relaxation time of supercooled liquids measured at variable temperatures and pressures. The proposed generalisation of the relaxation time could pave the way to a clearer connection between thermodynamic and dynamic parameters of a given system.

**RESULTS**

**VFT-like description of the dynamics of liquids and glasses**

We use fast scanning calorimetry to determine the heat capacity of glasses of indomethacin and toluene. We infer values of relaxation time at the onset devitrification temperature, $T_{on}$, by applying the known relationship $\tau_1\beta_1 = \tau_2\beta_2$[26]. A reference value of $\tau_1 = 100$ s, considered as the relaxation time of the glass at $T_{on}$ when heated at $\beta_1 = 0.167$ K/s[5,27], is employed, though we remark that slight variations on this value would yield equivalent conclusions. On the other hand, we also estimate the transformation time of each glass at the maximum of the transformation peak using the expression $t_{trans}(T_{max}) = \Delta T/\beta_m$, where $\Delta T$ is the width of the transformation peak and $\beta_m$ the mid value of the heating rate during the transformation. Further details about the calculation of the relaxation and transformation times from heat capacity data can be found in the methods section. As shown in the supplementary figure 3 both quantities yield comparable values. In the following we indistinctly use both measures to gauge the dynamics of the liquid and glassy states. Previous works have already considered this equivalence[28]. Further support of their likeness can be found in the Supplementary Information.

To quantify the stability of the glass we use the enthalpic limiting fictive temperature, $T'_f$, defined as the temperature at which the glass and the supercooled liquid have the same enthalpy[29]. At this temperature, the glass does not evolve thermodynamically. We remark that the measured values of limiting fictive temperature are independent of the heating rate of each calorimetric scan[15,30]. The choice of a convenient heating rate, in the range $0.0167 - 2 \cdot 10^4$ K/s, permits us to keep the system trapped in its initial glassy state along a larger temperature range, covering up to 75 K in temperature between the slowest and the fastest heating rates, while measuring the heat capacity during the transformation[15]. Figure 1 portrays data of both relaxation (open squares) and transformation times (closed squares) for three different glasses: (a) vapour-deposited indomethacin glasses grown at T = 266, 290, 300 and 310 K and a liquid-cooled glass, CG, cooled at -0.0167 K/s; (b) vapour-deposited toluene glasses grown at 111, 113 and 116 K in equilibrium with the liquid state and (c) liquid-cooled Au-based bulk metallic glasses aged to equilibrium at 373 and 383 K (data from ref.[22]). The relaxation times of the respective supercooled liquids are represented by triangles. The pink dashed line in Figure 1a represents Arrhenius behaviour and is included to better visualize the non-Arrhenius description of the high temperature data.

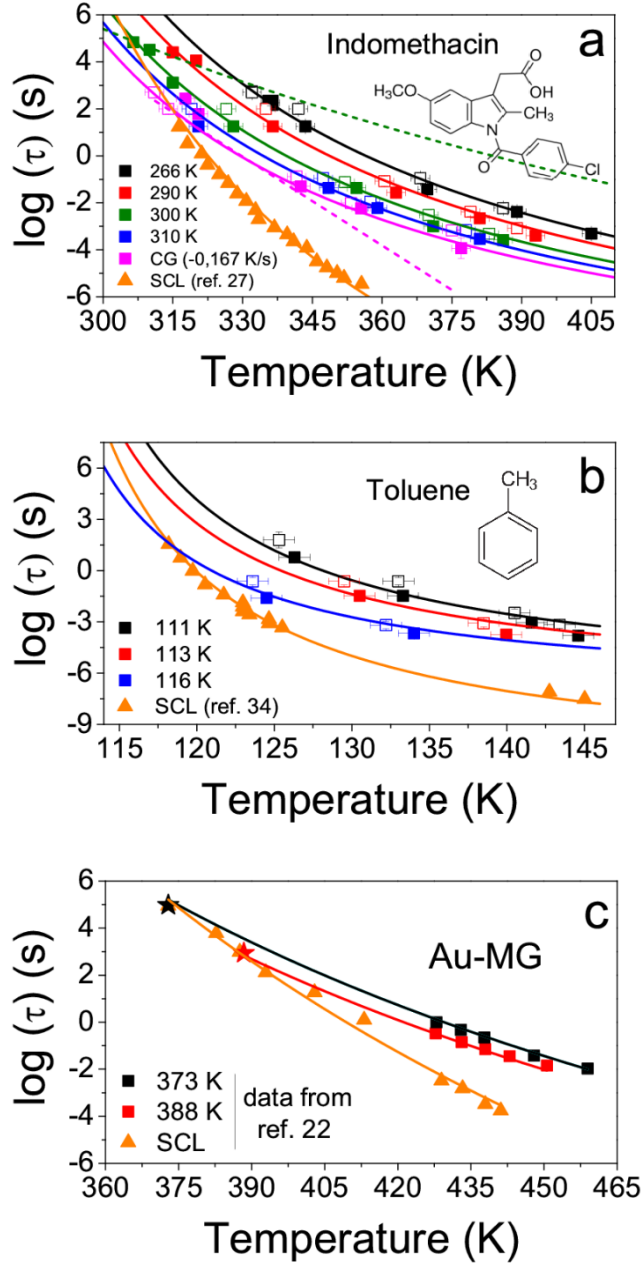

**Figure 1. Relaxation times (open symbols) and transformation times (filled symbols) derived from calorimetry experiments for three materials.** (a) IMC, (b) toluene and (c) Au-based bulk metallic glass (BMG) from ref.[22] with different stabilities and alpha relaxation times of their respective SCL (triangles). The temperatures highlighted as legends in the graphs correspond to deposition temperatures for IMC and toluene and to the aging temperature for the Au-based BMG. The stars in Fig. 1c are estimated points assuming that at $T'_f$ the transformation time of the glass equals the equilibrium relaxation time. The solid lines correspond to the best fit of the experimental points using equation (2). The fit parameters are presented in Table 1. The green dashed line in (a) corresponds to the glass relaxation time of a glass with $T'_f = 304$ K calculated with the Adam-Gibbs-Vogel (AGV) equation[17]. The pink dashed line in the same graph corresponds to an arbitrary Arrhenius curve, showing that the experimental data clearly exhibit super-Arrhenius behaviour. Error bars in relaxation time data calculated using the expression $\tau_2 = \tau_1 \beta_1 / \beta_2$ have been determined considering an uncertainty of $\pm 50$ s in $\tau_1$, and propagating it together with the uncertainty of $\pm 0.25 \beta_2$ in $\beta_2$. Error bars in transformation time data calculated using the expression $t_{trans}(T_{max}) = \Delta T / \beta_m$ have been determined by error propagation, considering an error of 1 K in $\Delta T$ and $0.25\beta_m$ in $\beta_m$. The uncertainty corresponding to the temperature axis is 2 K.

To search for a common description of the experimental data of Fig. 1 we propose a generalisation of equation (1) aimed at describing the dynamics of supercooled liquids and glasses with different thermal stability:

$$\tau_g = \tau_{g0} e^{\frac{\xi(T_f')T_0}{(T-T_0)}} \qquad (2)$$

where all the parameters have an analogous meaning as in equation (1). In this case, however, D has been substituted by a linear function of the limiting fictive temperature of the glass, $\xi(T_f',) = AT_f' + B$. In a supercooled liquid the fictive temperature $T_f = T$ at all temperatures, from the definition of $T_f$. We remark that for a given glass-former, IMC, toluene or BMG, the only non-shared parameter in the fittings is $T_f'$. That is, all curves have the same fitting parameters but different values of $T_f'$. The only exception is the limiting fictive temperature of the conventional IMC glass cooled at -0.0167 K/s, which for convenience is set to 315 K[31]. The resulting values of $T_f'$ that yield the simultaneous fitting of all glasses and the SCL are in reasonable agreement with the measured enthalpic limiting fictive temperature of each glass (Supplementary Table 2). Table 1 shows the values of the fitting parameters. The green dashed line in Figure 1a clearly highlights that in the probed temperature range, our experimental results differ from those predicted by the non-linear Adam-Gibbs-Vogel (AGV) equation[17] which has been often applied to understand out-of-equilibrium behaviour in a short temperature range around the conventional $T_g$[17].

Considering that equations (1) and (2) should be equivalent for a supercooled liquid, we derive the following equalities (see Supplementary Information):

$$D = AT_0 + B \qquad (3)$$

$$\tau_0 = \tau_{0g} e^{AT_0} \qquad (4)$$

The calculated values of D and $\tau_0$ are also shown in Table 1. Parameter D obtained for the supercooled liquid is related to the fragility, m, of the liquid[32]. Evaluation of m yields m = 54 for the Au-based BMG, considering a $T_g$ = 395 K ($\tau_\alpha = 100$ s), in fair agreement to the value measured by Wang et al., m = 49[22]. The obtained fragility value for IMC is 89, similar to that measured by Wojnarowska et al.[27] using dielectric spectroscopy, m = 83. In the case of toluene, we obtain a fragility m = 131. Kudlik et al.[33] reported a fragility parameter for toluene of m = 122, while from the VFT expression reported by Hatase et al.[34], m = 130. We note that, under this framework, the values of D and $\tau_0$ could be obtained from relaxation data corresponding uniquely to the glassy state. This is in accordance with some previous works claiming that the properties of the supercooled liquid may be embedded in the properties of their glasses[35,36].

**Table 1.** Parameters obtained by simultaneous fitting of the relaxation times for glasses with different stability and for the supercooled liquid using equation (2). $\tau_0$ and D have been calculated using equation (3) and (4) respectively.

|  | $T_0$ (K) | A (K$^{-1}$) | B | $\tau_{g0}$ (s) | $\tau_0$ (s) | D |
|---|---|---|---|---|---|---|
| **Indomethacin** | 230.54 | -0.106 | 44.93 | 2.69e-12 | 8.9e-23 | 20.55 |
| **Toluene** | 105.19 | -0.108 | 15.3 | 5.5e-8 | 7.08e-13 | 3.94 |
| **Au-BMG** | 129.45 | -0.222 | 203.45 | 3.98e-23 | 1.82e-35 | 174.75 |

**Superposition of relaxation times**

In the following we analyse the common description of the liquid and glassy state from another perspective. It has been shown that van der Waals' bonded liquids and polymers obey power-law density scaling[23–25], which means that the average relaxation time of the liquid is a function of $Tv^\gamma$, where $v(T, P) = 1/\rho$ is the specific volume and $\gamma$ is a material constant. The idea behind the scaling of relaxation times arises from the consideration that the local dynamics of liquids are governed by a generalised repulsive potential that scales with $\gamma$, under the assumption of spherical symmetry. This assumption is not strictly valid for interactions such as hydrogen bonds, although even in these cases the power-law scaling yields superposition of relaxation times as a function of T and v.[25] Although the scaling relationships were originally formulated for supercooled liquids, we extend them to glasses with different stabilities by introducing a dependence of the specific volume of the system on the limiting fictive temperature. In Fig. 2a we represent our relaxation data as a function of $1000\rho(T, T'_f)^\gamma/T$, where we set $\gamma = 6.53$. The detailed derivation of density values is given in the methods section.

Casalini et al.[25] derived the expression $\tau_\alpha(T, \rho) = F(Tv^\gamma)$ considering that the relaxation time is governed by the entropy of the system, $S_c$, as the AG model proposes, but using a generalised equation for $S_c$ that takes into account the influence of both temperature and, also, pressure (or, equivalently, changes in specific volume). In particular,

$$\tau(T, v) = \tau_0 \exp\left(\frac{C}{Tv^\gamma}\right)^\phi \quad (5)$$

where $\tau_0$ and $\phi$ are constants and $C = \left[\ln\left(\frac{100}{\tau_0}\right)\right]^{\frac{1}{\phi}} T_g v_g^\gamma$, with $T_g$ the conventional value of glass transition temperature for IMC, 315 K, $v_g$ the specific volume of a conventional glass at that temperature and $\gamma$ is the scaling parameter used in Fig. 2a. As in the case of the scaling relationship, we substitute the effect of pressure on specific volume for that of glass stability and express v as $v(T, T_f')$. The experimental data shown in Fig. 2b have been simultaneously fitted using equation (5), setting free the parameters $\tau_0$, $\phi$ and $\gamma$. The values of $v(T, T_f')$ are derived as indicated in the Methods section. The best fit is obtained with $\tau_0 = 2.26 \cdot 10^{-8}$ s, $\phi = 3.55$ and $\gamma = 6.53$. Alternatively, we can also infer the value of $\gamma$ from the slope of the $\log T_g$ vs $\log v_g$ curve, where $T_g$ and $v_g$ refer to the temperature and the specific volume of the system when the relaxation time equals 100 s, obtaining a value of 7 for the IMC glasses (see Supplementary Fig. 7).

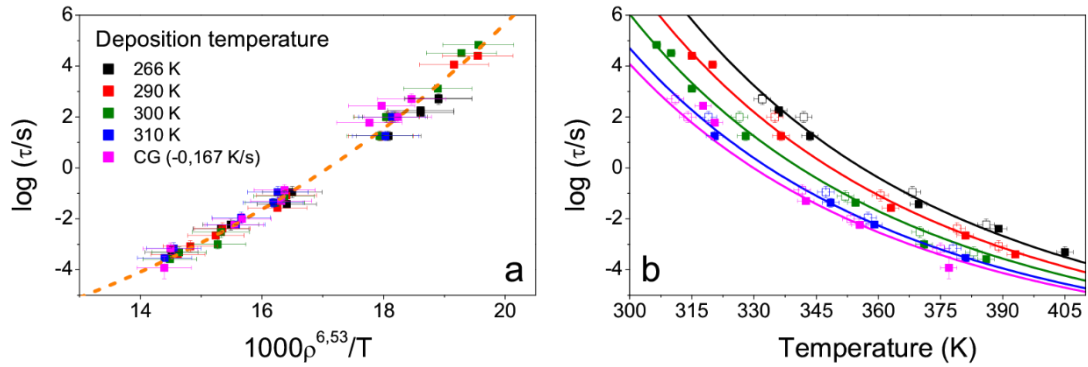

**Figure 2. Scaling relationship of the relaxation time of glasses of IMC with different stability and of the supercooled liquid.** (a) as a function of $1000\rho(T, T_f')^\gamma/T$, where $\gamma = 6.53$. The calculation of $\rho(T, T_f')$ is detailed in the methods section. (b) as a function of temperature. The continuous lines are the best fit of the experimental points using equation (5) and $\rho(T, T_f')$. The parameters $\tau_0$, $\phi$ and $\gamma_G$ are allowed to adjust freely. Error bars in the abscissa axis have been determined by error propagation, considering an uncertainty of 0.003 g/cm$^3$ in density and 2 K in temperature. Error bars in relaxation and transformation times have been determined as in Fig. 1.

The possibility to infer $\gamma$ from measurements at ambient pressure, in the liquid and glassy states, is promising. These findings support the idea that the dynamical behaviour of liquids and glasses can be explained and analysed under the same theoretical framework. Surprisingly, even the most stable glass ($T_{dep} = 266$ K), where hydrogen bonding between molecules are more abundant[37], is also reasonably well fitted by equation (5). This is compatible with the pressure dependence of the glass transition, $dT_g/dP$, evaluated for IMC by Wojnarowska et al.[27] The high value of $\lim_{P\to 0}(dT_g/dP)$ = 254 K/GPa, indicates that IMC could be regarded as a typical Van der Waals liquid.

**DISCUSSION**

We first focus on the potential role of the structure of the glass on the analysis of Fig. 1a. It is relatively well established that molecular packing anisotropy is a common characteristic of many vapour-deposited glasses[20,38,39]. Recent studies on thin film ultrastable glasses have shown that the transformation into the SCL proceeds through a heterogeneous mechanism starting at surfaces/interfaces and that the growth front velocity does not uniquely depend on the enthalpy content of the glass[15,40]. Our previous study concluded that the heterogeneous transformation of vapour-deposited thin film glasses of IMC could be divided into two families depending on the value of their birefringence, $\Delta n$[40]. Glasses with large birefringence ($> |0.02|$) exhibit much larger growth front velocities compared to glasses with small birefringence ($< |0.02|$). It is therefore worth interrogating whether anisotropy or molecular packing plays any role in the homogeneous transformation of the glass into the supercooled liquid. If this was the case, one should question the validity of equation (2) to simultaneously fit the liquid and glassy state, since this equation is a function of the enthalpy state of the glass, expressed through its limiting fictive temperature. Based on previous data[39], the glasses analysed here have $\Delta n \approx 0$, except those grown at 266 K with a low $\Delta n \approx 0.02$. We assume that the dynamics of the system during the bulk transformation is affected by the same parameters that affect the front transformation. Therefore, the simultaneous fit of the various glasses and the liquid state using a function of the enthalpy state of the glass is successful because those glasses behave as nearly isotropic from the point of view of the transformation into the SCL. In fact, attempts to include in Fig. 1a IMC glasses vapour-deposited at lower temperatures ($T_{dep} < 250$ K) and therefore with larger negative values of birefringence ($\Delta n < -0.02$) were not successful. It is important to note that the organic glasses analysed in the present work, those shown in Fig. 1a,b, are of bulk-type in the sense that their thickness is thick enough so they melt through a homogeneous process. In particular, the most stable IMC glasses exhibit homogeneous transformation for thicknesses above 900 nm[15], while for less stable glasses, the thickness threshold is lower[40].

It is interesting to note that glasses of two very different families, molecular and metallic, could be adjusted using equation (2). The liquid-cooled Au-based metallic glasses measured in ref.[22] and shown in Fig. 1c were aged for long times and equilibrated at the two temperatures of 373 and 383 K before being scanned up at fast heating rates using a Flash DSC. It is worth pointing out that beta relaxation processes are typically important in metallic glasses[41], and, in fact, short time aging of the Au-based glasses produced a simultaneous decrease of both $T_{on}$ and $T_f$, in clear contradiction with the equation (2). However, at the longer aging times needed for equilibration, the alpha relaxation time dominates over the beta relaxation and a decrease of $T_f$ is accompanied by an increase of $T_{on}$. It is however early to draw more general statements due to the scarcity of data in the high temperature regime. The popularisation of fast scanning methods will allow, in the years to come, to test the validity of VFT-type equations, such as equation (2), on a much larger number of materials. On the other hand, we are aware that a single fictive temperature value does not provide a unique description of the state of the glass[42–44]. However, our analysis suggests that a single enthalpic $T_f'$ offers a reasonable account of the dynamics of the glass in the medium-to-high temperature regime. We assume that the behaviour observed here is specific to glasses with a sufficiently narrow spatial distribution of inhomogeneities to allow for a single $T_f'$ description of the glass.

The fact that IMC glasses obey analogous density scaling relations as the supercooled liquid suggests that there are two relevant parameters controlling the dynamics in both the liquid and the glassy state: temperature and density. The scaling parameter obtained for glasses, $\gamma_{glass} = 6.53$, is, however, different to that of the supercooled liquid, $\gamma_{SCL} = 3.84$, obtained from reported PVT data[45] (see Supplementary Information for the derivation of this value). A common scaling exponent for all IMC glasses and the supercooled liquid could only be obtained with an unrealistic value of $\gamma = 9.1$, very far from the experimental value reported for the SCL. This may seem at odds with the common description of Figure 1. In the first section of this paper, we have shown that the relaxation data of IMC glasses and the supercooled liquid measured at atmospheric pressure could be simultaneously fitted using the same VFT-type expression, where the only variables were temperature and the limiting fictive temperature. However, the density of glasses and the supercooled liquid is not univocally determined by the fictive temperature of the system. In fact, the isobaric thermal expansion coefficient of IMC supercooled liquid at $T = 315$ K is $\alpha_{p,SCL} = 5.69 \cdot 10^{-4}$ K$^{-1}$, while, in the case of the IMC glass in equilibrium with the liquid at the same temperature is $\alpha_{p,SCL} = 1.32 \cdot 10^{-4}$ K$^{-1}$. Therefore, it is not surprising that the relaxation dynamics of glasses and their supercooled liquid can be simultaneously described using an expression with the limiting fictive temperature as variable parameter and not using the density.

The relation between the scaling exponent, $\gamma$, obtained from data fitting using equation 5 and the Grüneisen parameter

$$\gamma_G = \frac{\left(\frac{c_p}{c_v} - 1\right)}{T\alpha_p} \qquad (6)$$

is also a subject of intense research[23,25,46]. While at the origin these two parameters were considered to be equivalent, it was found significant discrepancy between them[25]. This discrepancy was recently solved by proposing that the energy distribution of the activation barrier for molecular rearrangements depends on the density of the system[46],

$$E_{barrier} \sim \left(\frac{\rho}{\rho_0}\right)^{\gamma_{EOS}} \quad (7)$$

where $\gamma_{EOS}$ is a constant.

Under this framework, the scaling exponent from equation 5 is reinterpreted as

$$\gamma = \frac{\gamma_{EOS}}{D} + \gamma_G \quad (8)$$

where D is the same as in equation 5. We saw before that $\gamma_{glass} = 6.53$ and $\gamma_{SCL} = 3.84$, while, from equation 6, $\gamma_{G,glass} = 0.79$ (glass with $T_f = 279$ K) and $\gamma_{G,SCL} = 2.25$. These values yield, according to eq. 8, $\gamma_{EOS,glass} = 20.38$ and $\gamma_{EOS,SCL} = 7.6$, meaning that the energy of the activation barriers is more sensitive to density changes in the case of the non-equilibrium glassy state than in the supercooled liquid.

**CONCLUSIONS**

In essence, we establish that the temperature dependence of the relaxation time for two organic and one metallic glass exhibit a super-Arrhenius behaviour in a medium-to-high temperature range. More importantly, generalised VFT-type equations that depend on the average limiting fictive temperature of the glass can be used to simultaneously describe the relaxation time of nearly-isotropic glasses with different stabilities and the supercooled liquid. The density scaling of glasses with different stability using an expression originally derived for supercooled liquids reinforce the analogy between the dynamic behaviour of glasses and liquids. We hope this work will help other researchers to establish closer connections between the liquid and glassy states of matter.

**METHODS**

**Growth and calorimetry measurements.**

IMC layers with thicknesses ranging from 600 nm to 2 µm were grown by thermal evaporation in a UHV chamber at $3\cdot 10^{-8}$ mbar, using an effusion cell (CREATEC) held at a constant temperature of around 440 Kelvin. IMC crystalline powder (99.9% purity) was acquired from Sigma-Aldrich and used as received. The evaporation rate, set at 0.15 nm/s, was monitored with a quartz microbalance (Sycon) located close to the substrate. Samples with different stabilities were produced by depositing them at different substrate temperatures, from 266 to 310 K. A liquid nitrogen cold trap was used to reduce the vapour pressure of certain contaminants, especially water. Conventional glasses have been produced by heating a deposited layer above their glass transition temperature, 315 K, and cooling them at a constant cooling rate of -10 K/min. The choice of thicknesses ensured that the main mechanism of the transformation into the supercooled liquid was homogeneous through the entire sample and not heterogeneous as occurs in thinner films.

In order to study the transformation kinetics of the deposited glasses along a wide temperature range, different calorimetric techniques and methodologies were applied.

(1) In the high temperature range ($\tau$ below $10^{-2}$ s), quasi adiabatic fast-scanning calorimetry is employed[13,15,16]. Fast heating rates (from $10^3$ to $10^5$ K/s), raise the glass transition temperature by tenths of degrees. The samples are deposited onto a membrane-based calorimetric cell. A shadow mask placed between the calorimetric cell and the vapour-flux assures that the material is deposited within the sensing area of the device (1 mm$^2$). Prior to the experiments, a 200 nm aluminium film is deposited onto the sensing area of the membrane to improve the temperature distribution. A model[13] is applied in order to obtain heat capacity data from the raw voltage data obtained from the measurement.

(2) To measure the transformation kinetics in the medium-to-high temperature range ($\tau$ between 1 and $10^{-2}$ s), we apply a non-constant intensity to the same nanocalorimetric cell, increasing its value with time, with the possibility of reaching constant but intermediate heating rates, ranging from 10 to $10^3$ K/s. At these heating rates, the measurements are not strictly adiabatic and, therefore, thermal losses between the sample and the environment are present. From the apparent heat capacity we extract the onset temperature and the width of the transformation peak.

(3) Differential Scanning Calorimetry with a Perkin Elmer DSC7 is used to measure the transformation kinetics in the low temperature range ($\tau$ between $10^2$ and 1 s). We deposit 1.5 µm thick samples onto aluminium foil, which is subsequently folded and introduced into a DSC aluminium pan. The time between the extraction of the sample from the deposition chamber and the placement of the pan into the DSC cell was reduced at maximum to avoid water absorption into the glass.

(4) The transformation times at the lowest temperature range ($\tau$ above $10^2$ s) were determined by isothermal experiments. In the case of samples with intermediate stability (with $T_f > 280$ K), in-situ isotherms were performed in order to avoid water absorption during the process. In those measurements 1.5 µm thick layers are deposited onto the calorimetric cell and kept at a given temperature (annealing temperature). After time t, a calorimetric scan at low heating rate is performed to determine the onset temperature of the annealed sample. From the Cp curve we can know whether the sample has been transformed or not. The represented value of transformation time corresponds to the mean between the larger annealing time of the non-transformed samples and the shorter annealing time of a completely transformed sample. In the case of ultrastable glasses ($T_{dep}$ = 266 K), the power output of the DSC was registered during an isotherm at the temperature of interest, following the sample preparation method described previously in point (3).

In all cases except for the isothermal experiments, the deposition temperature was controlled by the device itself, feeding it with a non-variable value of intensity during the deposition, reaching a constant temperature. In the case of the isothermal experiments, the deposition temperature is controlled by means of heating resistances and a Pt100 sensor attached to the socket where the measuring device is placed.

The limiting fictive temperature of glasses grown at different deposition temperatures is measured by integrating the specific heat data obtained from slow heating rates measurements performed in the DSC and from ultra-fast heating rate measurements performed by quasi-adiabatic fast-scanning nanocalorimetry in thin layers. The details of the procedure have been described elsewhere[15]. In the case of intermediate heating rates, the non-adiabatic conditions of the experiment preclude the proper evaluation of reliable values of limiting fictive temperature.

**Analysis of heat capacity data: Derivation of relaxation and transformation times.**

Once the heat capacity is derived from the raw data, we perform the following analysis to obtain the values of transformation and relaxation times. In the first case, we employ the expression $\tau_1 \beta_1 = \tau_2 \beta_2$ to calculate the relaxation time, $\tau_2$, of a glass of a given stability at the onset temperature of the transformation when heated at a given rate, $\beta_2$, considering as reference value of relaxation time $\tau_1 = 100$ s when the heating rate is $\beta_1 = 0.167$ K/s (Supplementary Fig. 1a). In the second case, we consider the temperature at the maximum of the transformation peak for each glass measured at each heating rate. The transformation time corresponding to this temperature is calculated as $t_{trans}(T_{max}) = \Delta T/\beta_m$, where $\Delta T$ is the peak amplitude at its base and $\beta_m$ is the mid value of heating rate (Supplementary Fig. 1b). The methodology is tested by comparing the transformation times obtained with this approach to those measured through isothermal measurements at specific temperatures, (details given in supplementary information). The width of the transformation peaks evaluated at a given temperature ($T_{max}$) remains approximately constant for glasses of different stabilities. This fact, together with the observation by Talansky et al.[47] that the distribution of transformation times in a vapour-deposited glass of methyl-m-toluate was around 25%, permits us to infer that the potential variation of this parameter among the different glasses, if any, is below our experimental uncertainty in the evaluation of $\Delta T$ and will not affect our conclusions.

**Calculation of density as a function of stability and temperature**

The density of the conventional IMC glass at ambient conditions is 1.31 g/cm$^3$ [48]. The density of indomethacin glasses with different stability is calculated from the density variations reported by Dalal et al.[39], measured at 293 K (Supplementary Fig. 5). The variation of density with temperature has been calculated from the reported thermal expansion coefficients, $\alpha_{UG} = 1.39 \cdot 10^{-4}$ K$^{-1}$, $\alpha_{CG}(T_g = 309$ K$) = 1.33 \cdot 10^{-4}$ K$^{-1}$ and $\alpha_{SCL} = 5.69 \cdot 10^{-4}$ K$^{-1}$, [39]. For intermediate stabilities, a linear interpolation between these values has been performed. The reference density of supercooled IMC has been chosen to be equal to the density of the conventional glass, $T_f' = 315$ K, at $T_{ref} = 315$ K.

$$\rho(T, T_f) = \frac{\rho_0(T_f', T_{ref})}{1 + \alpha_T(T_f')(T - T_{ref})} \quad (6)$$

Different values of conventional IMC glass density have been reported[49]. However, it should be noted that while the choice of a different reference value of density shifts the curves towards lower or higher values of $Tv^\gamma$, it does not appreciably change the scaling factor.

## ACKNOWLEDGEMENTS

This research was supported by MINECO through Projects MAT2013-40986-P and MAT2014-57866- REDT. We thank discussions with K. Ngai, M. Ediger and J.L. Tamarit.

## AUTHORS' CONTRIBUTION

CRT and MGS conceived the idea, CRT and JRR performed the experiments, CRT and MGS analysed the data, JRV and MGS supervised the work, CRT and JRV, with the help of all authors, wrote the paper. All authors contribute to the discussion of the results.

## COMPETING FINANCIAL INTERESTS

The authors declare no competing financial interests.

# Supplementary Information.

# Relaxation dynamics of glasses along a wide stability and temperature range

C. Rodríguez-Tinoco, J. Ràfols-Ribé, M. González-Silveira, J. Rodríguez-Viejo

Physics Department, Universitat Autònoma de Barcelona, 08193 Bellaterra, Spain.

## Derivation of relaxation and transformation times.

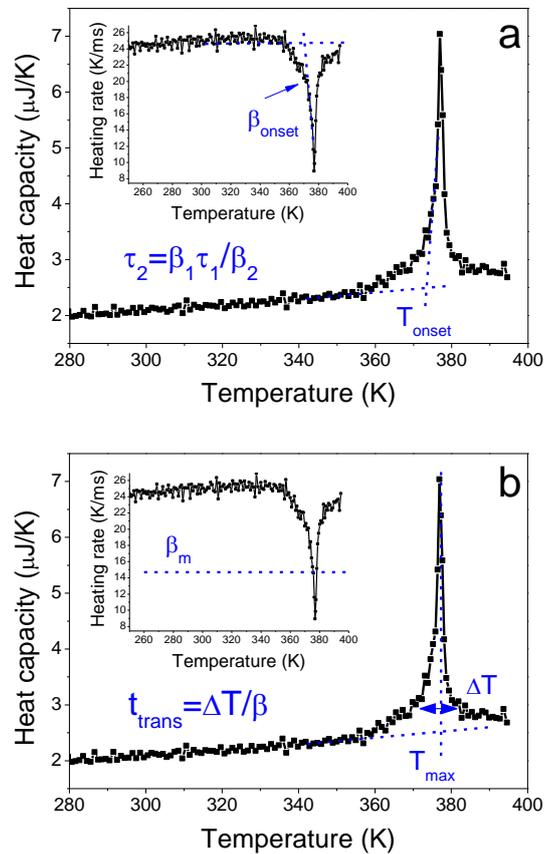

**Supplementary Figure 1. Schema of the calculation of relaxation and transformation times from heat capacity data.** a) in the first approach, we use the expression $\tau_1 \beta_1 = \tau_2 \beta_2$ to obtain the value of glass relaxation time from the heating rate of the experiment, assigning this value to the onset temperature. b) in the second approach, we calculate the transformation time from the width of the transformation peak and the midpoint value of the heating rate, assigning it to the temperature at the maximum of the transformation peak.

**Comparison between procedures to determine relaxation and transformation times**

In the supplementary Fig. 2 we show the comparison among different methods to infer the value of relaxation time of a glass with a particular stability and heated at a given rate. In the Supplementary Fig. 2a, we compare the transformation time calculated, on one hand, using the expression referred in the main text, $t_{trans}(T_{max}) = \Delta T/\beta$, and, on the other, the transformation time directly measured from an isotherm measure at the same temperature performed in a conventional DSC. The transformation time is considered to be the time elapsed from the beginning of the isotherm process to the moment at which the power output of the DSC is constant. The transformation time obtained from the two methods, 182 and 150 seconds respectively, are fairly comparable.

On the other hand, we can derive the relaxation time of a glass at the onset of the transformation measured at a given rate from the well-known expression $\tau_1 \beta_1 = \tau_2 \beta_2$, taking as reference values $\tau_1 = 100\ s$ and $\beta_1 = 0.167\ K/s$. In particular, for an ultrastable glass measured at $\beta_2 = 0.033\ K/s$, $\tau_2 = 506\ s$. We can compare this value to the one obtained by performing an isothermal measurement at the same temperature ($T_{on} = 332\ K$), 550 s, in fair agreement with the previous result (Supplementary Fig. 2b).

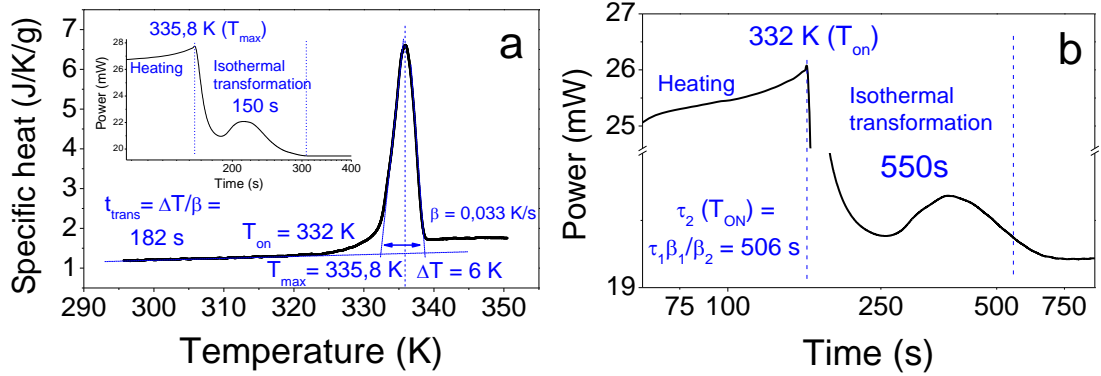

**Supplementary Figure 2. Comparison between procedures to determine relaxation times.** a) DSC scan of an IMC glass deposited at $T_{dep}$ = 266 K, heated at 0.033 K/s. From the width of the peak and the heating rate, the transformation time is inferred as indicated and assigned to the temperature where the maximum of the peak appears. In the inset, a DSC isotherm on an equivalent sample performed at the temperature of the maximum of the peak is shown. From that measurement, we find the transformation time of the sample at that temperature. From the two measurements, we find that both methodologies are approximately equivalent. b) a DSC isotherm on an equivalent sample performed at the onset temperature of the transformation is shown. The transformation time is fairly equivalent to the relaxation time inferred using the expression $\tau_1\beta_1 = \tau_2\beta_2$, as explained in the text.

In Supplementary Fig. 3 we plot the relaxation time (a) and transformation time (b) calculated for glasses with different stability and measured at different heating rates, together with the structural relaxation time published for IMC supercooled liquid. The fitting of these data using equation (2) yields similar curves, as seen in Supplementary table 1. Also, the limiting fictive temperature values obtained by fitting the experimental points shown in Supplementary Fig. 3 a and b, and also obtained from the combined data shown in Fig. 1a, yields similar results, as seen in Supplementary Table 2.

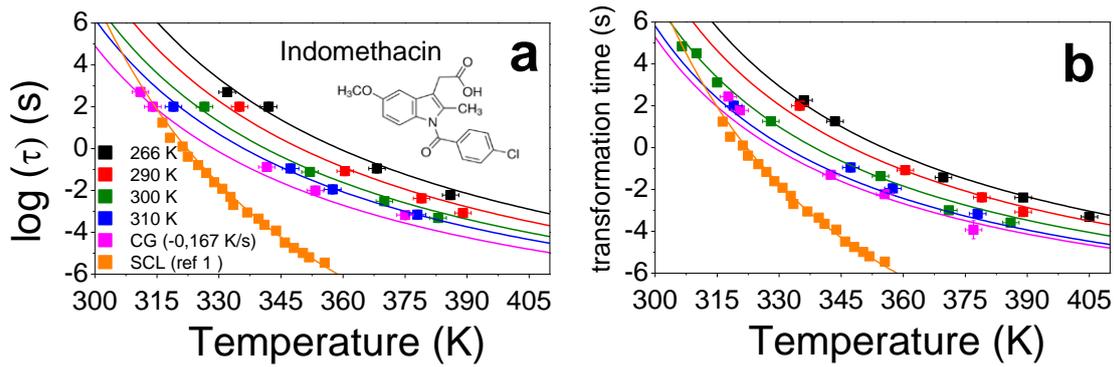

**Supplementary Figure 3. Comparison between the adjustment of relaxation and transformation times with equation (2).** Relaxation (a) and transformation (b) times of IMC glasses with different stability measured at different heating rates and structural relaxation time of supercooled IMC liquid from ref[1]. The fits correspond to equation (2). The fit parameters are shown in supplementary table 1 and 2.

From these observations, we can infer that: i) the relaxation time of glasses, calculated using the expression, $\tau_1\beta_1 = \tau_2\beta_2$, and that of the supercooled liquid can be described using the same empirical relationship, and ii) the similitude between the fitting parameters obtained using the expression above for the relaxation times and those determined from the transformation times, is indicative of the similarity between the two concepts, at least in the experimental conditions under which our experiments were performed. For all this, we consider both measurements as representative of the same magnitude, the relaxation time of the glass.

The agreement between the nominal limiting fictive temperature of the measured glasses and $T_f'$ obtained by fitting the experimental data using equation (5) can also be seen from the data shown in Supplementary Table 2.

**Supplementary Table 1**. Fitting values using equation (2) of the main text. $\tau$ refers to relaxation times calculated from $\tau_1\beta_1 = \tau_2\beta_2$ assuming $\tau_1 = 100$ s for $\beta_1 = 0.167$ K/s, $t_{trans}$ refers to transformation times. $\tau + t_{trans}$ refers to the fitting using all data. The later are used in the main text.

| Indomethacin | | $T_0$ (K) | A (K$^{-1}$) | B | $\tau_{g0}$ (s) | $\tau_0$ (s) | D |
|---|---|---|---|---|---|---|---|
| Fitted data | $\tau$ | 228.59 | -0.107 | 45.62 | 1.91e-12 | 5.75e-23 | 21.16 |
| | $t_{trans}$ | 229.86 | -0.105 | 44.8 | 2.57e-12 | 1.07e-22 | 20.66 |
| | $\tau + t_{trans}$ | 230.54 | -0.106 | 44.93 | 2.69e-12 | 8.9e-23 | 20.55 |

**Supplementary Table 2.** Values of limiting fictive temperature for IMC glasses resulting from the various fittings using equation (2) and equation (5) compared to nominal values, $T_{f,\ nominal}$, obtained by integration of the heat capacity curves.

| Indomethacin | | | | | | |
|---|---|---|---|---|---|---|
| $T_{dep}$ (K) | | | 266 | 290 | 300 | 310 |
| $T_f'$ fit (K) (using eq. (2)) | Fitted data | $\tau$ | 284.5±1.5 | 293.3±1.2 | 303.7±1 | 309.3±1.2 |
| | | $t_{trans}$ | 284.9±1.4 | 293.6±1.5 | 303.9±1 | 309.4±1.1 |
| | | $\tau + t_{trans}$ | 285.2±1.4 | 293.8±1.4 | 304±1 | 309.6±1.1 |
| $T_{f,\ nominal}'$ (K) | | | 279±2.5 | 289±2.5 | 301±2.5 | 311.5±2.5 |
| **Toluene** | | | | | | |
| $T_{dep}$ (K) | | | 111 | 113 | 116 | |
| $T_f'$ fit (K) | | $\tau + t_{trans}$ | 107.2±1.3 | 111.4±1.1 | 118.3±0.8 | |
| $T_{f,\ nominal}'$ (K) | | | 111±2.5 | 113±2.5 | 116±2.5 | |

# Representation of relaxation data as a function of the inverse of temperature

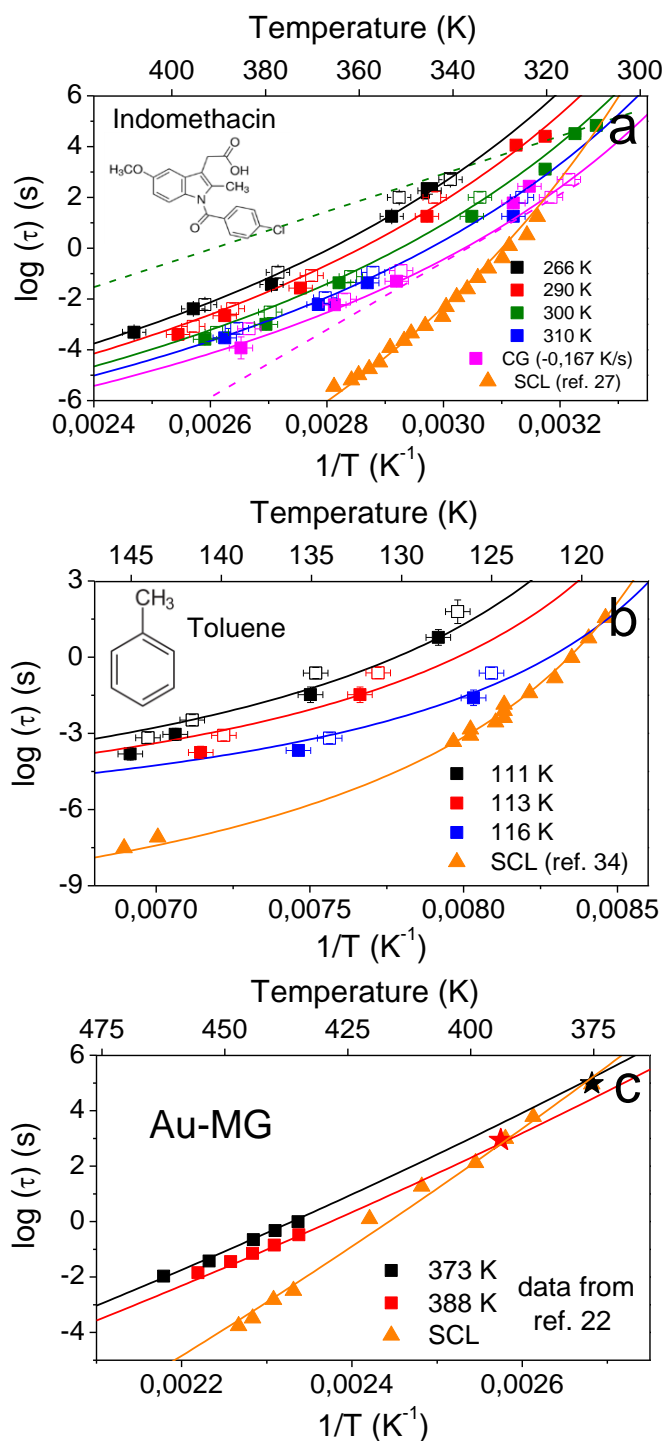

**Supplementary figure 4.** Representation of the relaxation data shown in Figure 1 as a function of 1/T, to better visualize the non-Arrhenius dependence of the glass relaxation time in a sufficiently extended temperature range. References from the main text.

### Derivation of D and $\tau_0$ from equation (2)

In supercooled liquid, $T_f = T$ at all the temperatures. In this case, equation (1) and (2) should coincide. Therefore,

$$\tau_{g0} e^{\frac{\xi(T)T_0}{(T-T_0)}} = \tau_0 e^{\frac{DT_0}{(T-T_0)}}$$

Taking natural logarithms and isolating, $\xi(T)$, we obtain that,

$$\xi(T) = -\frac{1}{T_0}\ln\left(\frac{\tau_{0g}}{\tau_0}\right)T + \left(D + \ln\left(\frac{\tau_{0g}}{\tau_0}\right)\right)$$

Assuming the expression, $\xi = AT_f + B$, we see that,

$$A = -\frac{1}{T_0}\ln\left(\frac{\tau_{0g}}{\tau_0}\right)$$

$$B = \left(D + \ln\left(\frac{\tau_{0g}}{\tau_0}\right)\right)$$

From where we obtain the equations (3) and (4) for D and $\tau_0$.

## Calculation of density as a function of stability and temperature

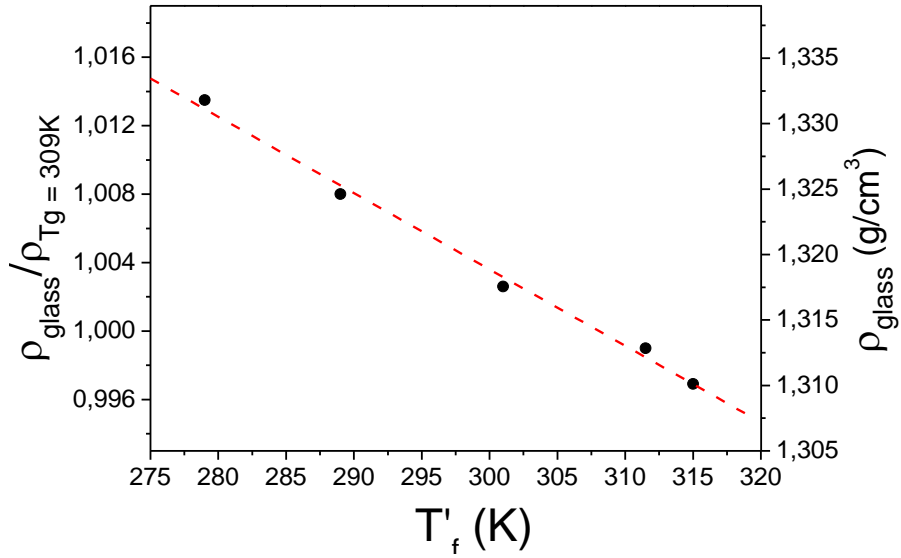

**Supplementary Figure 5. Density of IMC glasses with different stability at 293 K extracted from ref**[2]. In the left axis, data is represented with respect to the density of a glass cooled at 1 K/min ($T'_f$ = 309 K), as shown in the reference. Right axis has been stablished after the consideration that, for $T'_f$ = 315 K (conventional glass), the density is 1.31 g/cm$^3$. Transformation from $T_{dep}$ data to $T'_f$ has been performed according to the relationship between the two quantities, as shown in ref[3].

**Supplementary table 3.** Values of density and thermal expansion coefficients used for each glass and for the supercooled liquid in equation (5) to construct the Supplementary Fig. 4.

| $T'_{f,nominal}$ (K) | $\rho_0 \left(\frac{g}{cm^3}\right)$ at 293 K | $\alpha_T (K^{-1}) \times 10^{-4}$ |
|---|---|---|
| **279** | 1.332 | 1.39 |
| **289** | 1.325 | 1.374 |
| **301** | 1.318 | 1.354 |
| **311.5** | 1.313 | 1.33 |
| **315 (SC -10 K/min)** | 1.31 | 1.32 |
| **SCL** | 1.307 (at 315 K) | 5.69 |

**Derivation of scaling parameter from relaxation time data of supercooled IMC liquid**

In Supplementary Figure 6a, we plot the relaxation time of supercooled indomethacin liquid as a function of temperature and at different isobars, measured by Wojnarowska et al.[1]. In order to obtain the scaling parameter, we fit this experimental data using the equation 5 from the main text. The specific volume as a function of temperature at different isobars is extrapolated from the PVT data reported by Adrjanowicz et al.[4]. The best fit is obtained with $\tau_0 = 4.68 \cdot 10^{-10}$ s, $\phi = 4.78$ and $\gamma = 3.84$.

In Supplementary figure 6b we plot the $\log T_g$ vs $\log v_g$, where $T_g$ and $v_g$ refer to the temperature and specific volume of the system at the transition from liquid to glass, obtained from the reported PVT data[4]. From the slope of this curve, the scaling parameter can also be found, according to[5]

$$\log T_g = A - \gamma \log v_g$$

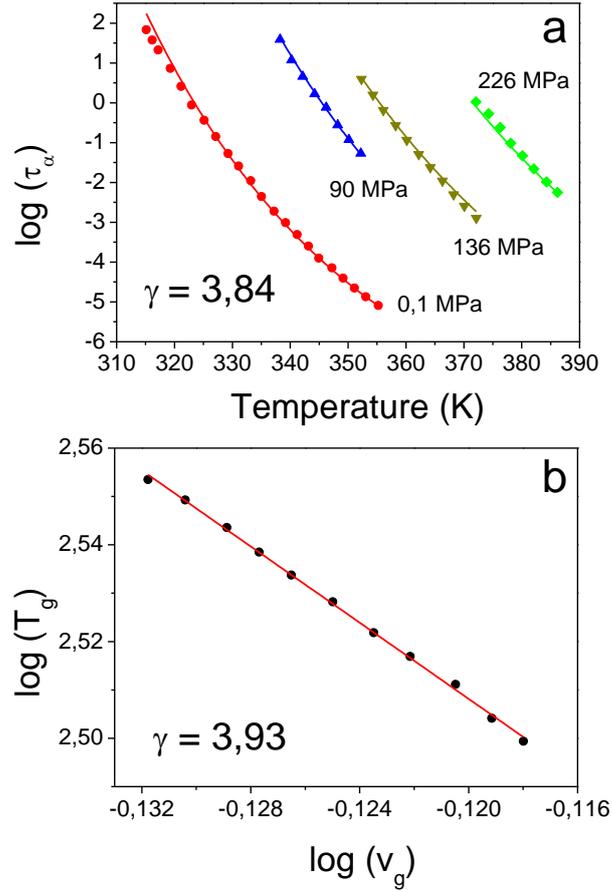

**Supplementary figure 6.** a) Relaxation time of supercooled IMC at different isobars. The curves are fitted using equation 5 from the main text, yielding a scaling parameter of 3.84. b) logT$_g$ vs logv$_g$, where T$_g$ and v$_g$ refer to the temperature and specific volume of the system at the transition from liquid to glass. All experimental data have been extracted from reported results[1,4].

**Alternative calculation of scaling parameter from glass relaxation time data**

From the relaxation data plotted in Figure 2b in the main text, we obtain T$_g$ as the temperature at which the relaxation time of the glass equals 100 s. At this temperature, the specific volume of the glass is $v_g$. From the slope of the representation shown in Supplementary figure 7, we can calculate the scaling factor, according to

$$\log T_g = A - \gamma \log v_g$$

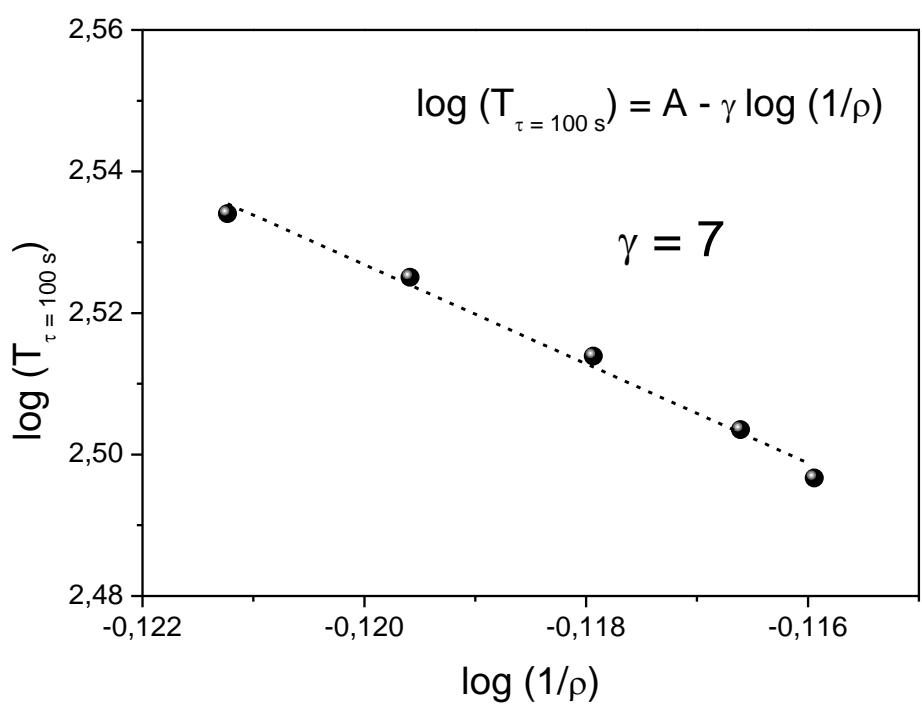

**Supplementary figure 7**. $\log T_g$ vs $\log v_g$, where $T_g$ and $v_g$ refer to the temperature and specific volume of the system when the relaxation time equals 100 s. The slope of the fitted curve corresponds to the scaling parameter.

We note the similitude between the scaling parameter found from the fitting of the relaxation time using equation 5 in the main text ($\gamma_{glass} = 6.53$) and using the approach in Supplementary figure 7 ($\gamma_{glass} = 7$).

**SUPPLEMENTARY REFERENCES**